\documentclass{appolb}
\usepackage{graphicx}

\begin{document}
\title{The Matrix Element Method at next-to-leading order accuracy %
\thanks{Presented by T. Martini at the XXXIX International Conference of Theoretical
Physics `Matter to the Deepest', Ustro\'n, Poland, September 13th-18th, 2015.}%
}
\author{Till Martini, Peter Uwer
\address{Humboldt-Universit\"{a}t zu Berlin, Institut f\"{u}r Physik\\
  Ph\"{a}nomenologie der Elementarteilchenphysik}
}
\maketitle
\begin{abstract}
The Matrix Element Method (MEM) has proven beneficial to make maximal use of the information available in experimental data. However, so far it has mostly been used in Born approximation only.
In this paper we discuss an extension to NLO accuracy. As a prerequisite we present an efficient method to calculate event weights for jet events at NLO accuracy.
As illustration and proof of concept we apply the method to the extraction of the top-quark mass in $e^+e^-$ annihilation.
We observe significant differences when moving from LO to NLO which may be relevant for the interpretation of top-quark mass measurements at hadron colliders relying on the MEM.\\
\newline
DOI: {10.5506/APhysPolB.46.2143}\hfill HU-EP-15/57
\end{abstract}
\section{Jet event weight}
In experiments, jets are obtained from hadronic momenta which are reconstructed from tracks and energy depositions in the detectors. In theory, those jets are modelled by clustering the partonic final states one usually ends up with in calculations:
$
p_1,\dots, p_N\rightarrow J_1,\dots, J_n.
$
In what follows we will refer to all observed (jet-)momenta $J_1,\dots, J_n$ from a collision $P_A+P_B\rightarrow J_1,\dots, J_n+X$ as `jet event'. A weight for this event can be defined by interpreting the differential cross section as a probabilty density to measure a specific event:
$
\rho=\frac{d\sigma_{AB\rightarrow n}}{dJ_1\dots dJ_n}.
$
With a differential jet cross section, one can calculate more inclusive observables, generate unweighted events according to $\rho$  or use it in likelihood analysis methods like the Matrix Element Method.
\section{Matrix Element Method}
\subsection{Leading order}
The MEM \cite{Kondo:1988yd} can be used to extract model parameters $\Omega$ from data by maximizing a likelihood (e.g. for a given set of jet events $\vec{x_i}=(J_1,...,J_n)_i$)

\[
\mathcal{L}^{\textrm{\scriptsize Born}}(\Omega)=\prod\limits_{i}\frac{1}{\sigma^{\textrm{\scriptsize B}}(\Omega)}\int d\vec{y}\frac{d\sigma^{\textrm{\scriptsize B}}(\Omega)}{d\vec{y}}\!\!\!\underbrace{W(\vec{x_i},\vec{y})}_{\textrm{here: }=\delta(\vec{x}-\vec{y})\!\!\!\!}=\prod\limits_{i}\frac{1}{\sigma^{\textrm{\scriptsize B}}(\Omega)}\frac{d\sigma^{\textrm{\scriptsize B}}(\Omega)}{d\vec{x_i}}
\]
which is proportional to the differential cross section. 
 Maximizing with respect to $\Omega$ yields an estimator for $\widehat{\Omega}$:
$
\mathcal{L}^{\textrm{\scriptsize Born}}(\widehat{\Omega})={\sup_{\Omega}}{\;\mathcal{L}^{\textrm{\scriptsize Born}}(\Omega)}.
$
Since all information in the event is used in the matrix element, this method is believed to provide the most efficient estimator. Pioneered at the Tevatron (e.g. \cite{Abazov:2004cs,Abulencia:2006mi}), the MEM is widely used today (e.g. \cite{Artoisenet:2013vfa,Khiem:2015ofa}). Automation of the MEM has been studied in \cite{Artoisenet:2010cn}. However, one caveat of the above formulation is its limitation to LO accuracy. A first attempt towards NLO has been made in \cite{Alwall:2010cq} studying the effect of QCD radiation. In \cite{Soper:2014rya}, the hard matrix element and a parton shower is used to discriminate signal \textit{versus} background. An NLO extension for events with uncolored final states has already been presented in \cite{Campbell:2012cz}. The hadronic production of jets is investigated in \cite{Campbell:2013uha}. NLO and LO jets are mapped by means of a boost along the beam axis to balance the transverse momentum.

\subsection{Next-to-leading order likelihood for jet events}
When writing an ansatz for the likelihood at NLO (for a given set of events with $n$ jets) as a sum of the Born and virtual (B+V) part and the real (R) part (with unresolved additional radiation) 
\[
{ \mathcal{L}^{\textrm{\scriptsize NLO}}(\Omega)=\prod\limits_{i}\frac{1}{\sigma^{\textrm{\scriptsize NLO}}_{\textrm{\scriptsize $n$-jet}}(\Omega)}\left({\frac{d\sigma^{\textrm{\scriptsize B+V}}_{n\rightarrow\textrm{\scriptsize $n$-jet}}(\Omega)}{dJ_1\dots dJ_n}}+{\frac{d\sigma^{\textrm{\scriptsize R}}_{n+1\rightarrow\textrm{\scriptsize $n$-jet}}(\Omega)}{dJ_1\dots dJ_n}}\right)\!\!\Bigg|_{{\textrm{event } i}}}
\]
one faces three problems: 
\begin{enumerate}
\item Since both contributions are separately infrared (IR) divergent, a point-wise cancellation within phase space must be ensured. So both contributions have to be evaluated for the same jet momenta. 
\item In the real contribution, $n+1$ partons are clustered to $n$ jets: $J_i=\widetilde{J}_i(p_1,\dots,p_{n+1})$. This introduces $\delta$-functions $\delta(J_i-\widetilde{J}_i(p_1,\dots,p_{n+1}))$ in the phase space integration which render any numerical integration useless. If the real phase space factorizes in terms of an $n$-jet phase space and the unresolved configurations:
$
dR_{n+1}(p_1,...,p_{n+1})=dR_n(\widetilde{J}_1,...,\widetilde{J}_n)dR_{\textrm{\scriptsize unres}}(\Phi)
$,
these $\delta$-functions will be fulfilled by construction:
$
dR_{n+1}(p_1,...,p_{n+1})\delta(\widetilde{J}_i-{J_i})=dR_{\textrm{\scriptsize unres}}(\Phi)\big|_{\widetilde{J}_i=J_i}.
$
This factorisation serves as an inversion of the jet algorithm since $dR_{\textrm{\scriptsize unres}}(\Phi)$ generates only partonic configurations that result in the given jet event. 
\item In the Born and virtual contribution, $n$ partons form $n$ jets: $J_i=p_i$. To evaluate the Born and virtual matrix elements for the jet momenta ($J_i=\widetilde{J}_i$), the clustered jets have to be on-shell: $\widetilde{J}^2_i=m^2_i$ and respect momentum conservation: $\widetilde{J}_1+\dots+ \widetilde{J}_n=p_1+\dots+p_{n+1}$ at the same time. This is not possible with $2\rightarrow  1$ clustering prescriptions. 
\end{enumerate}
Employing instead $3\rightarrow  2$ clustering prescriptions (inspired by the Catani-Seymour dipole subtraction method \cite{Catani:1996vz,Catani:2002hc}) allows to meet these requirements enabling the cancellation of the IR divergences and factorization of the real phase space. Using this modified jet algorithm allows to define an event weight (differential jet cross section) at NLO accuracy
\[
\frac{d\sigma^{\textrm{\scriptsize NLO}}_{\textrm{\scriptsize $n$-jet}}(\Omega)}{dJ_1\dots dJ_n}
={\frac{d\sigma^{\textrm{\scriptsize B+V}}(\Omega)}{dJ_1\dots dJ_n}}
+{{\int\!\!dR_{\textrm{\scriptsize unres}}(\Phi)}}{\frac{d\sigma^{\textrm{\scriptsize R}}(\Omega)}{dp_1\dots dp_{n+1}}}.
\]
The mutual cancellation of the IR divergencies has to be carried out by a suitable subtraction method (e.g. phase space slicing \cite{Harris:2001sx}).
\section{Validation \& application}
To valididate our approach, we reproduce jet distributions calculated with an conventional parton level Monte-Carlo generator (MC) using the $3\rightarrow 2$ jet algorithm. Two sample processes which cover all essential aspects are studied: Drell--Yan $pp\rightarrow e^+e^-$ (initial state radiation) and top-quark pair production at a lepton collider $e^+e^-\rightarrow t\bar{t}$ (final state radiation with massive particles). For calculational details, see \cite{Harris:2001sx,Brandenburg:1998xw}. We veto on additional jet emission since we are only interested in the case when there is a recombination in the real contribution.
\subsection{Phase space generation}
\begin{figure}[htb]
\centerline{%
\includegraphics[width=9.6cm]{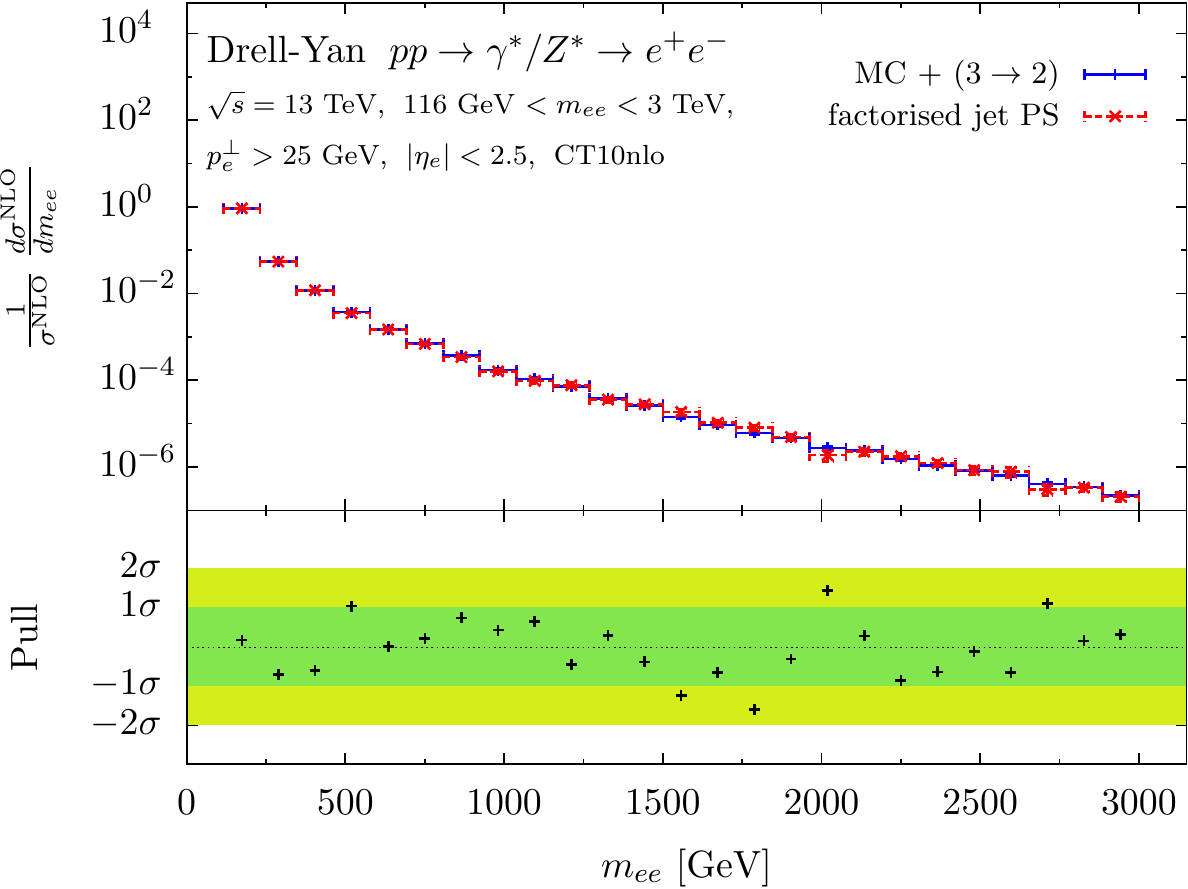}}
\caption{Validation of the phase space generation (Drell--Yan)}
\label{Fig:psvaldy}
\end{figure}
As an example Fig. \ref{Fig:psvaldy} shows the invariant mass distribution of the lepton pair from Drell--Yan obtained with the parton level MC (blue solid) and the factorized jet phase space as outlined above (red dashed). The lower part of the plot shows the difference in terms of the statistical error. We find perfect agreement between the two distributions. Comparisons of other observables, also for top-quark pair production yield similar conclusions and are presented in \cite{Martini:2015}.
\subsection{Generation of unweighted events}
\begin{figure}[htb]
\centerline{%
\includegraphics[width=9.6cm]{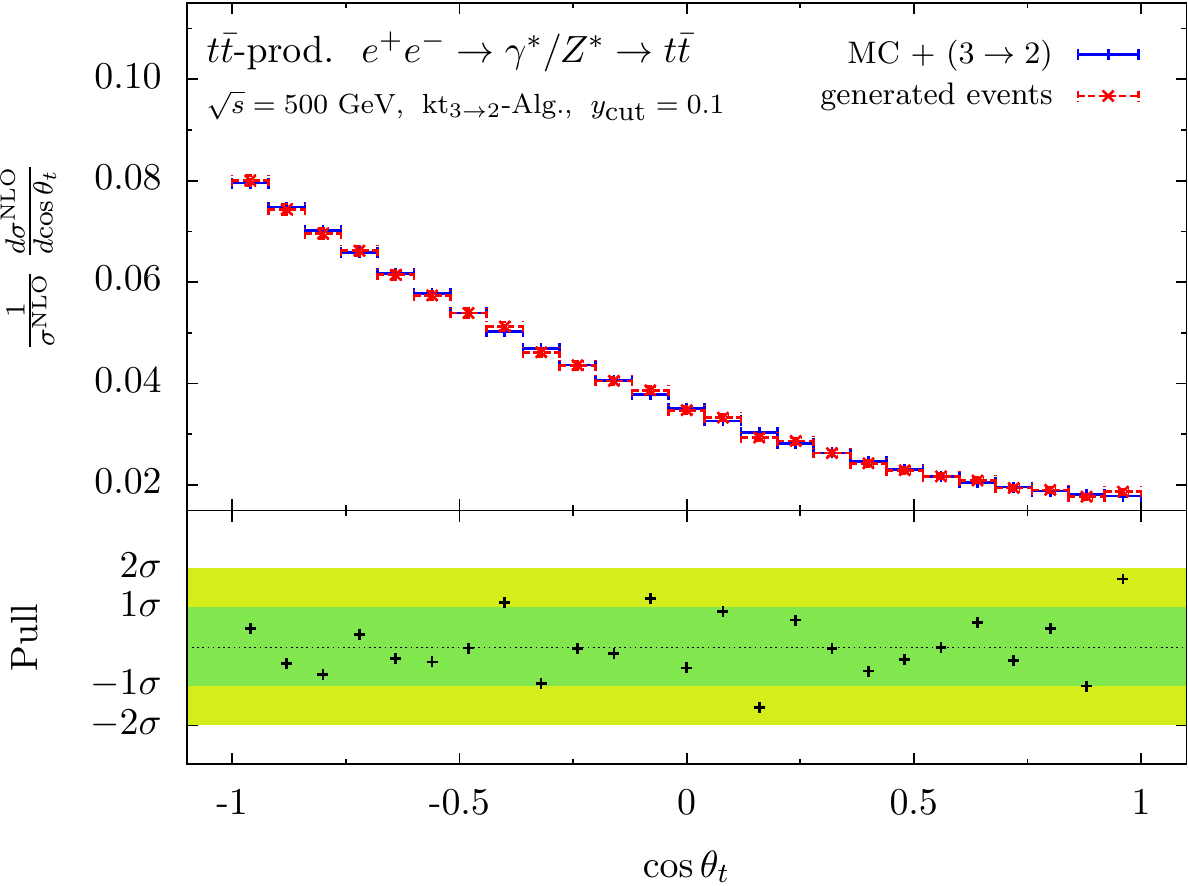}}
\caption{Validation of the generation of unweighted NLO top-quark pair events}
\label{Fig:uwevaltt}
\end{figure}
With the jet event weight at NLO, we can generate unweighted jet events at NLO using a simple acceptance/rejection algorithm. As an example, Fig. \ref{Fig:uwevaltt} shows histogrammed $\cos\theta_t$ of these events (red dashed). Again we find perfect agreement when comparing with the distribution obtained from the conventional parton level MC (blue solid). Comparisons of the other observables yield similar conclusions (see \cite{Martini:2015}).
 \subsection{Matrix Element method at next-to-leading order accuracy}
\begin{figure}[!htb]
\centerline{%
\includegraphics[width=9.6cm]{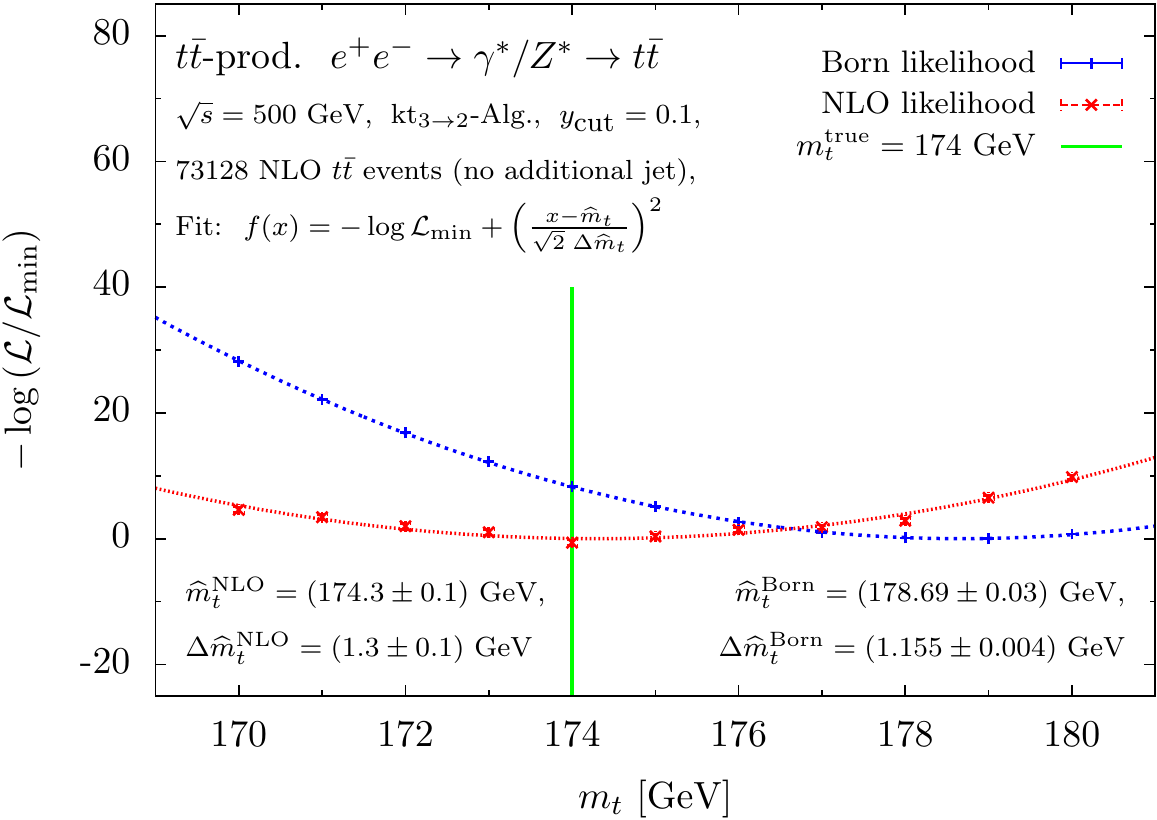}}
\caption{NLO and Born Log-likelihood as a function of $m_t$}
\label{Fig:mem}
\end{figure}

Treating the sample of unweighted jet events which were generated with an input value of the top-quark mass $m^{\textrm{\scriptsize true}}_t=174$ GeV as a toy experiment, where jet angles $\vec{x}_i=(\cos\theta_t,\phi_t,\cos\theta_{\bar{t}},\phi_{\bar{t}})_i$ were measured, we can apply the MEM at NLO to the sample of NLO $t\bar{t}$ events
\[
\mathcal{L}^{\textrm{\scriptsize NLO}}(m_t)=\prod\limits_{i}^N\mathcal{L}^{\textrm{\scriptsize NLO}}(\vec{x}_i|m_t)=\left(\frac{\beta_t}{32\pi^2\sigma^{\textrm{\scriptsize NLO}}_{t\bar{t}}(m_t)}\right)^N\prod\limits_{i}^N\frac{d\sigma^{\textrm{\scriptsize NLO}}_{t\bar{t}}(m_t)}{dJ_{t}\; dJ_{\bar{t}}}\bigg|_{\textrm{event } i}.
\]
Fig. \ref{Fig:mem} shows the negative logarithm of the likelihood (`Log-Likelihood') as a function of $m_t$ at Born (solid/blue) and NLO (dashed/red) accuracy. The estimator $\widehat{m}_t\pm\Delta\widehat{m}_t$ is extracted by a fit (dotted lines). Extracting $\widehat{m}^{\textrm{\textrm{\scriptsize NLO}}}_t$ from the NLO events with the NLO likelihood perfectly reproduces $m^{\textrm{\scriptsize true}}_t$, while the extraction with the Born likelihood yields a value for $\widehat{m}^{\textrm{\textrm{\scriptsize Born}}}_t$ which differs from $m^{\textrm{\scriptsize true}}_t$ by $4$ GeV. It is no surprise that using the wrong likelihood might result in a biased estimator. However, the size of the effect is remarkable, considering that NLO corrections to distributions for this process only amount to a few percent. Finally, it should be noted that the renormalization scheme is well-defined in the MEM at NLO, allowing a less ambigious interpretation of the extracted parameters. For more details on
MEM at NLO, we refer the reader to \cite{Martini:2015}.


\begin{thebibliography}{9}
\bibitem{Kondo:1988yd}
K.~Kondo.
\newblock {Dynamical Likelihood Method for Reconstruction of Events With
  Missing Momentum. 1: Method and Toy Models}.
\newblock {\em J.Phys.Soc.Jap.}, 57:4126--4140, 1988.

\bibitem{Abazov:2004cs}
V.M. Abazov et~al.
\newblock {A precision measurement of the mass of the top quark}.
\newblock {\em Nature}, 429:638--642, 2004, arXiv:hep-ex/0406031.

\bibitem{Abulencia:2006mi}
A.~Abulencia et~al.
\newblock {Top quark mass measurement from dilepton events at CDF II with the
  matrix-element method}.
\newblock {\em Phys.Rev.}, D74:032009, 2006, arXiv:hep-ex/0605118.

\bibitem{Artoisenet:2013vfa}
P.~Artoisenet, P.~de~Aquino, F.~Maltoni, and O.~Mattelaer.
\newblock {Unravelling $t\overline{t}h$ via the Matrix Element Method}.
\newblock {\em Phys.Rev.Lett.}, 111(9):091802, 2013, arXiv:1304.6414 [hep-ph].

\bibitem{Khiem:2015ofa}
P.H.~Khiem, E.~Kou, Y.~Kurihara, and F.~Le Diberder.
\newblock {Probing New Physics using top quark polarization in the $e^+e^-
  \rightarrow t \bar{t}$ process at future Linear Colliders}.
\newblock 2015, arXiv:1503.04247 [hep-ph].

\bibitem{Artoisenet:2010cn}
P.~Artoisenet, V.~Lemaitre, F.~Maltoni, and O.~Mattelaer.
\newblock {Automation of the matrix element reweighting method}.
\newblock {\em JHEP}, 1012:068, 2010, arXiv:1007.3300 [hep-ph].

\bibitem{Alwall:2010cq}
J.~Alwall, A.~Freitas, and O.~Mattelaer.
\newblock {The Matrix Element Method and QCD Radiation}.
\newblock {\em Phys.Rev.}, D83:074010, 2011, arXiv:1010.2263 [hep-ph].

\bibitem{Soper:2014rya}
D.E.~Soper and M.~Spannowsky.
\newblock {Finding physics signals with event deconstruction}.
\newblock {\em Phys.Rev.}, D89(9):094005, 2014, arXiv:1402.1189 [hep-ph].

\bibitem{Campbell:2012cz}
J.M.~Campbell, W.T.~Giele, and C.~Williams.
\newblock {The Matrix Element Method at Next-to-Leading Order}.
\newblock {\em JHEP}, 1211:043, 2012, arXiv:1204.4424 [hep-ph].

\bibitem{Campbell:2013uha}
C.~Williams, J.M.~Campbell, and W.T.~Giele.
\newblock {Event-by-event weighting at next-to-leading order}.
\newblock {\em PoS}, RADCOR2013:037, 2013, arXiv:1311.5811 [hep-ph].

\bibitem{Catani:1996vz}
S.~Catani and M.H.~Seymour.
\newblock {A General algorithm for calculating jet cross-sections in NLO QCD}.
\newblock {\em Nucl.Phys.}, B485:291--419, 1997, arXiv:hep-ph/9605323.

\bibitem{Catani:2002hc}
S.~Catani, S.~Dittmaier, M.H.~Seymour, and Z.~Trocsanyi.
\newblock {The Dipole formalism for next-to-leading order QCD calculations with
  massive partons}.
\newblock {\em Nucl.Phys.}, B627:189--265, 2002, arXiv:hep-ph/0201036.

\bibitem{Harris:2001sx}
B.W.~Harris and J.F.~Owens.
\newblock {The Two cutoff phase space slicing method}.
\newblock {\em Phys.Rev.}, D65:094032, 2002, arXiv:hep-ph/0102128.

\bibitem{Brandenburg:1998xw}
A.~Brandenburg, M.~Flesch, and P.~Uwer.
\newblock {The Spin density matrix of top quark pairs produced in electron -
  positron annihilation including QCD radiative corrections}.
\newblock {\em Phys.Rev.}, D59:014001, 1999, arXiv:hep-ph/9806306.

\bibitem{Martini:2015}
T.~Martini and P.~Uwer.
\newblock Extending the matrix element method beyond the born approximation:
  calculating event weights at next-to-leading order accuracy.
\newblock {\em Journal of High Energy Physics}, 2015(9), 2015, arXiv:1506.08798 [hep-ph].

\end{thebibliography}
\end{document}